\documentclass{iopjournal}

\usepackage{array}
\usepackage{placeins}
\usepackage{booktabs}
\usepackage{amsmath}
\usepackage{cleveref}
 
\begin{document}

\articletype{Paper} 

\title{Design of Microlens Arrays for Thermal Imaging with Spintronic Poisson Bolometers}

\author{Ziyi Yang$^{1}$\orcid{0009-0006-3175-5387}, Leif Bauer$^{1}$\orcid{0000-0002-7137-3379} and Zubin Jacob$^{1,*}$\orcid{0009-0007-5856-641X}}

\affil{$^1$Elmore Family School of Electrical and Computer Engineering, Purdue University, USA}\\
\email{zjacob@purdue.edu}


\begin{abstract}
Infrared (IR) detectors are widely used due to their ability to sense thermal radiation. Recently, a room-temperature infrared detector known as the spintronic Poisson bolometer was introduced. While offering fast digital readout, its sensitivity is currently limited by its small photosensitive area and array fill factor. Beyond this specific detector, many emerging detector architectures also require substantial in-pixel electronics or engineering tradeoffs, which can reduce the fill factor and degrade optical coupling efficiency. In this work, we present design guidelines for spherical plano-convex microlens arrays that enhance light collection in spintronic Poisson bolometer arrays in the mid-wave infrared (MWIR). Guided by the simulations, we fabricate a microlens array sample to demonstrate that the chosen geometrical parameters are experimentally attainable and compatible with the fabrication process. We then systematically explore the scaling laws between sensor active size, microlens geometries, and the resulting light collection metrics, yielding practical design rules that are broadly relevant to MWIR detector arrays with limited active area. A radiometric-stochastic model is used to quantify the resulting sensitivity improvements for the spintronic Poisson bolometer. Our work is the first systematic simulation of a microlens design with spintronic Poisson bolometer arrays, bridging microphotonics, spintronics, and thermal imaging.
\end{abstract}


\section{Introduction}

Infrared (IR) detectors have a wide range of applications in various fields due to their ability to detect thermal radiation, such as astronomy, biophotonics, defense, environmental monitoring, and autonomous vehicles \cite{Bao2023, Lahiri2012, Kaushal2017}. Recently, a room-temperature infrared detector called the spintronic Poisson bolometer (SPB) was introduced \cite{bauer2025exploiting, mousa2026ultra}. Unlike conventional analog infrared detectors that produce continuous electrical signals with Gaussian noise, the Poisson bolometer operates in a probabilistic regime governed by Poissonian counting statistics. In this framework, both the signal and the associated noise arise from discrete switching events, with incident radiation modulating the average transition rate independently of the photon statistics of the source \cite{yang2025optical}. This digital detection mechanism offers a fundamentally different approach to discriminating signal from noise and has the potential to relax conventional sensitivity constraints \cite{fossum2016quanta}. 

\begin{figure}[t!]
\centering
\includegraphics[width= 0.9\linewidth]{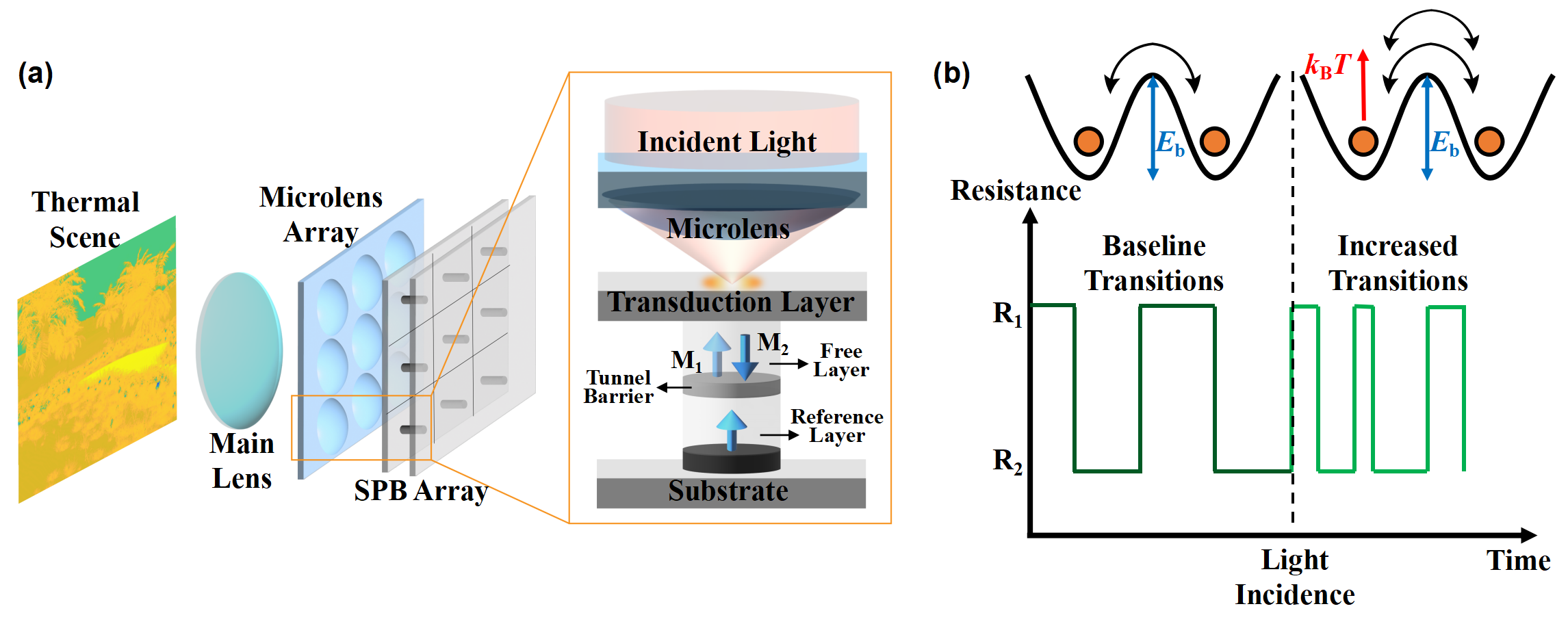}

\caption{(a) Schematic of the imaging system and spintronic Poisson bolometer device structure. Thermal infrared radiation emitted by the scene is focused by the main lens onto the microlens array, which is placed at the image plane of the main lens. The radiation is then focused by the microlenses onto the spintronic Poisson bolometer array, with an enlarged optically active area and enhanced optical coupling. The device exploits thermally activated transitions between two metastable magnetization states, $M_1$ and $M_2$, to produce a digital response under infrared illumination. (b) Operating mechanism of a spintronic Poisson bolometer. Top: Energy landscape of the free layer. The relaxation times of the two states are proportional to $exp(E_\text{b}/k_\text{B}T)$, where $E_\text{b}$ is the energy barrier between the two states, $k_\text{B}$ is the Boltzmann constant, and $T$ is the temperature of the free layer \cite{bauer2025exploiting}. When no light is incident, the temperature of the free layer is low, and the rate of the thermally activated transitions is low. When light is incident, the temperature increases, resulting in an increased rate of transitions. Bottom: Device resistance change. By applying a bias voltage to the device, this resistance change can be read out. The increased transitions signal the detection of light incidence. The number of transitions in a time window follows a Poisson distribution.}
\label{fig1}
\end{figure}

In \cite{bauer2025exploiting} and \cite{mousa2026ultra}, the Poisson bolometers were realized using engineered spintronic structures, namely a stochastic magnetic tunnel junction (MTJ), as shown in \autoref{fig1} (a). The device exploits thermally activated transitions between two metastable magnetization states, $M_1$ and $M_2$, to produce a digital response under infrared illumination (\autoref{fig1} (b)). The authors reported a 148 mK noise-equivalent differential temperature (NEDT) in the mid-wave infrared (MWIR) (3-5 $\mu m$) in \cite{mousa2026ultra}. However, the size of the MTJ pillar used in their spintronic Poisson bolometer cannot exceed several hundred nanometers due to the necessity of a small energy barrier $E_\text{b}$ that enables high-speed stochastic magnetization flips at room temperature \cite{bauer2025exploiting}. Although this nanoscale footprint enables high pixel density and potential array scaling, the limited active area, i.e., photosensitive area, also restricts the detector's sensitivity, especially under low-light conditions, such as those found in remote sensing and imaging of rapidly evolving scenes \cite{rashman2020terrestrial, landmann2023high}.

At the array level, this constraint on the active area can be quantified using the fill factor (FF) metric, defined as the ratio of the photosensitive area to the total pixel area \cite{bruschini2023challenges}. A higher fill factor indicates that a larger portion of each pixel actively detects light, leading to greater light collection efficiency. While mature MWIR imaging platforms, such as detector arrays based on Type-II superlattices (T2SL) photodiodes \cite{jasik2025demonstration}, mercury cadmium telluride (MCT/HgCdTe) photodiodes \cite{destefanis2007advanced}, and vanadium oxide (VO$_x$) microbolometers \cite{celik2020640}, can achieve high optical fill factors ($>70\%$), many emerging architectures require substantial in-pixel electronics \cite{hampel202464, taylor2026mid}, or engineering tradeoffs \cite{ramos2023optical}, which can reduce the geometric fill factor and degrade optical coupling efficiency.

A conventional approach to increasing the effective fill factor is to integrate detector arrays with a microlens array (MLA) \cite{bruschini2023challenges, kang2022substrate, pan2025metalens}. Three common planar microlenses include spherical microlenses, Fresnel zone plate (FZP) microlenses, and metalenses \cite{dilhan2020planar}. Among these, spherical microlenses are frequently utilized due to their straightforward design and geometric simplicity. Current literature exploring spherical MWIR microlens arrays predominantly focuses on case-specific optimizations \cite{allen2016increasing, jin2022light, lee2020design}. Typically, these studies employ rigorous computational methods, such as Finite-Difference Time-Domain (FDTD) \cite{allen2016increasing, jin2022light, lee2020design} or Rigorous Coupled-Wave Analysis (RCWA) \cite{pan2025design}, to simulate focusing efficiency and optical transmission. Although individual geometric parameters such as lens sag and radius of curvature are often optimized, generalized parametric studies that map the scaling laws between sensor active size, microlens diameter, lens sag, and light-collection performance remain limited.

In this paper, we show a parametric design of aluminum oxide-based (Al$_2$O$_3$) spherical microlens arrays for MWIR detectors based on FDTD simulations. Guided by these simulations, we fabricate a microlens array sample to demonstrate that the proposed geometrical parameters are compatible with practical fabrication. We further show through system-level modeling that microlens arrays can theoretically improve the imaging performance of the recently developed spintronic Poisson bolometer arrays by decreasing NEDT. Although microlens arrays are a well-established technology, their targeted use to engineer optical coupling for SPBs introduces a meaningful pathway to system-level enhancement for this emerging detector modality. Beyond the specific detector demonstrated here, this study also provides a systematic MWIR design framework that relates microlens geometry and detector active area to focal behavior and optical concentration, yielding practical design rules that are relevant to MWIR detector arrays with limited active area, and to our knowledge, are not commonly presented in a generalized form in the literature.

\section{Results}

\begin{figure}[t!]
\centering

\includegraphics[width= 1\linewidth]{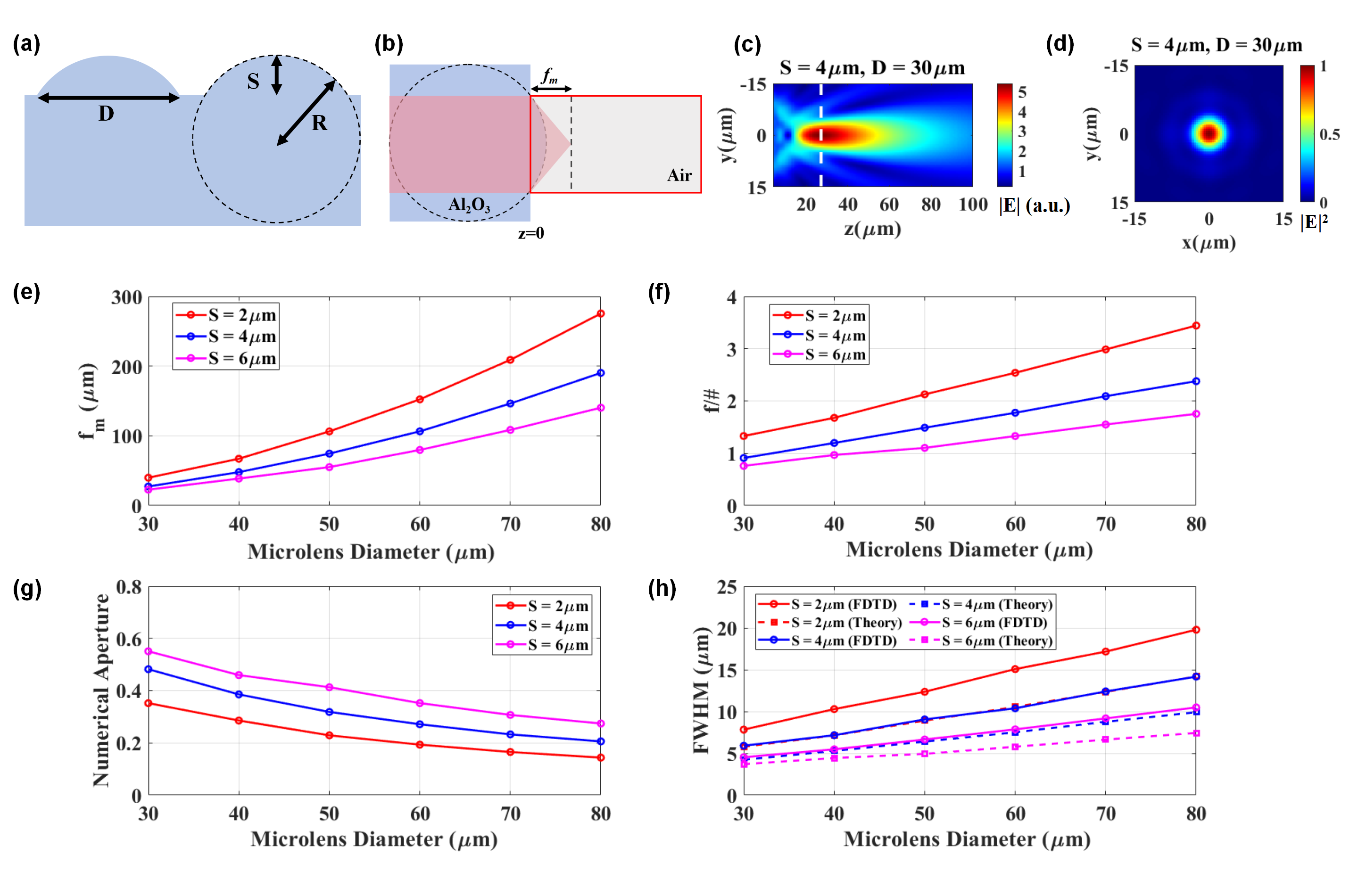}

\caption{FDTD simulation of Al$_2$O$_3$ spherical microlenses. (a) Plano-convex spherical microlens geometry. \( R \) is the radius of curvature, \( S \) is the lens sag, and \( D \) is the diameter of the microlens. (b) Configuration for Figure 2(c). The red box corresponds to the simulation region in Figure 2(c). (c) Longitudinal E-field distribution of a microlens with 4 $\mu m$ lens sag and 30 $\mu m$ diameter. The white dashed line marks the focal length. (d) Normalized E-field intensity at the focus. (e) Mechanical focal length (MFL) $f_m$ versus microlens diameter for different lens sags. (f) f-number versus microlens diameter for different lens sags. (g) Numerical aperture (NA) versus microlens diameter for different lens sags. (h) Spot size, characterized by FWHM, versus microlens diameter for different lens sags.}
\label{fig1.5}
\end{figure}

We begin by designing a spherical plano-convex microlens array for the MWIR. The basic parameters of a spherical plano-convex microlens are shown in \autoref{fig1.5} (a), where \( R \) is the radius of curvature, \( S \) is the lens sag, and \( D \) is the diameter of the microlens. We choose aluminum oxide (Al$_2$O$_3$) as the microlens material due to its high MWIR transmission (80\%) and compatibility with standard microlens fabrication techniques \cite{park2001refractive, sung2001high}. We performed finite-difference time-domain (FDTD) simulations of electromagnetic wave propagation through the microlens with linearly polarized plane wave illumination using Lumerical. Detailed settings of the FDTD simulation are included in Section 5.1.

\autoref{fig1.5} (b) shows the configuration of the simulation. The curved lens surface faces towards the right, and the light is incident from the left. The z=0 plane represents the flat base of the microlens. \autoref{fig1.5} (c) shows the E-field distribution, from which we extract the location of the focus. Then we record the normalized E-field intensity at the focus (\autoref{fig1.5} (d)). By repeating this simulation for different lens diameters and sags, we obtained mechanical focal length (MFL) $f_m$, f-number $f/\# =  \frac{f}{D}$, numerical aperture (NA), and spot size characterized by full width at half maximum (FWHM), as shown by \autoref{fig1.5} (e)-(h). 

Here, the mechanical focal length $f_m$ (the white dashed line in Fig. 2(c) and the curves in Fig. 2(e)) is extracted by finding out the maximum value of E-field $|E|$ along the z-axis at (x, y) = (0, 0) in Fig. 2(c) (same coordinate where the intensity reaches its maximum). $f_m$ is the total physical distance from the flat base of the microlens (the rim of the microlens) to the plane of maximum intensity. Unlike the standard effective focal length (EFL), which is measured from the vertex of the curved surface, here we approximated the focal length using the flat interface due to wafer integration considerations. A detailed explanation of this mechanical focal length can be found in Section 5.3. 

The numerical aperture in Fig. 2(g) is calculated by \cite{BornWolf1999}:

\begin{equation}
NA = n sin\varphi = sin\varphi = \frac{D/2}{\sqrt{f_m^2+(D/2)^2}}
\end{equation}

\noindent The theoretical FWHM spot size in \autoref{fig1.5} (h) is calculated using \cite{masters2020superresolution}:

\begin{equation}
  FWHM = 0.51\frac{\lambda}{NA}
\end{equation}

To verify the practical feasibility of these simulated geometries, we fabricated a series of aluminum oxide microlens arrays using the thermal reflow method \cite{nussbaum1997design}. First, a thickness of 2 $\mu m$ photoresist is spin-coated on a 1-inch diameter circular Al$_2$O$_3$ substrate. Next, arrays of circles with diameters ranging from 20 $\mu m$ to 34 $\mu m$ are patterned by photolithography. After standard resist development, arrays of resist cylinders are formed. The sample is then heated at 100$^\circ\mathrm{C}$ for 6 minutes on a hot plate, and the resist cylinders melt into semi-spheres. Finally, resist semispheres are used as masks to transfer their shape to the Al$_2$O$_3$ substrate by reactive ion etching (RIE).

\begin{figure}[ht!]
\centering

\includegraphics[width= 0.9\linewidth]{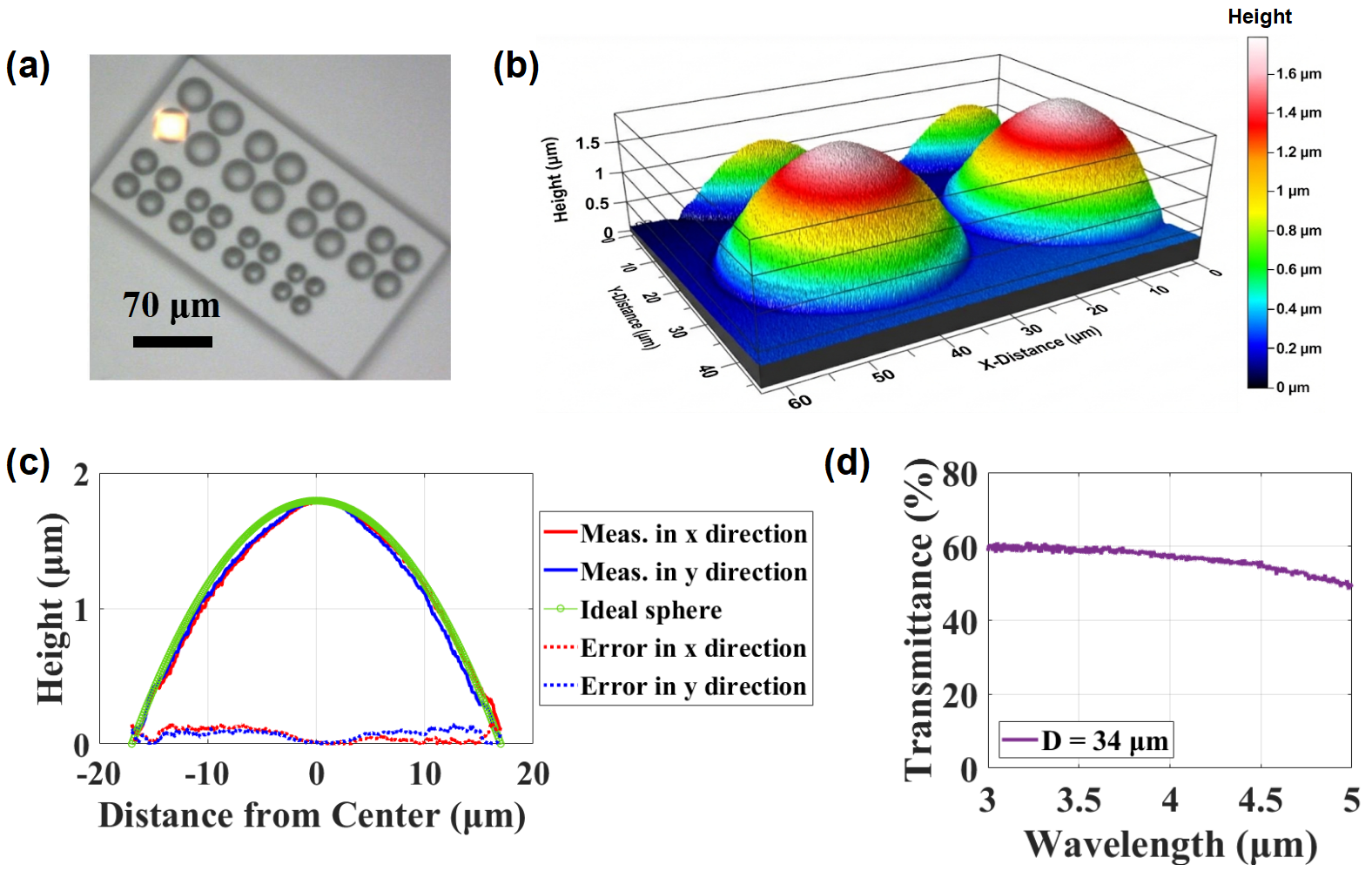}

\caption{Al$_2$O$_3$ microlens array fabrication and characterization of a 34 $\mu m$ diameter microlens. (a) Microscopic image of the MLA sample. The bright spot is the alignment beam of the FTIR spectroscopy. (b) Profilometry of the microlens with 34 $\mu m$ diameter. (c) X and Y direction surface profile compared to a perfect spherical shape. (d) Transmission spectra measured by FTIR spectroscopy in the MWIR.}
\label{fab}
\end{figure}

\autoref{fab} (a) shows a microscopic image of the fabricated microlens arrays. The sample consists of a series of 2 × 2 microlens arrays with microlens diameters ranging from 20 to 34 $\mu m$. \autoref{fab} (b) shows the measured surface profile of a 34 $\mu m$ microlens array using an optical profilometer (Bruker Contour GT-K). The actual lens sag decreased from the design value of 2 $\mu m$ to 1.79 $\mu m$, commonly expected due to the etch selectivity between the mask and the substrate during reactive ion etching (RIE) transfer \cite{nussbaum1997design, eisner1996transferring}. This can be compensated for by increasing the thickness of the photoresist mask or fine tuning the etch rate \cite{nussbaum1997design}. We compare the fabricated microlens profile to an ideal spherical shape in \autoref{fab} (c). The average error between the measured profile and the ideal spherical profile is 0.063 $\mu m$ ($\approx0.0158\lambda$). This error lies well within the subwavelength regime and is generally considered sufficiently small for microlens fabrication \cite{BornWolf1999, eisner1996transferring}. Therefore, the fabricated microlenses are expected to be suitable for most practical imaging applications \cite{roeder2018injection}. In \autoref{fab} (d), we measure transmission spectra of the fabricated microlens arrays in MWIR using Fourier Transform Infrared (FTIR) spectroscopy. The transmission maintains around 60\% across MWIR, indicating promising potential for practical applications \cite{zhang2021fabrication, li2022empowering, zhang2018solid}.

Next, we use the results from our FDTD simulations to evaluate the effect of MWIR microlens arrays on improving detector performance. We begin by evaluating the light-collection performance of a microlens array using two metrics. The first metric is collection efficiency (CE), also known as the focus efficiency, which is defined as \cite{kang2022substrate}:

\begin{equation}
  CE =  \frac{P_{\text{active}}}{P_{\text{pixel}}}
\end{equation}

\noindent where $P_{\text{active}}$ is the optical power incident on sensor active area and $P_{\text{pixel}}$ is the optical power incident on pixel area. The power incident on the sensor active area is calculated by integrating the intensity profile at the focus \cite{kang2022substrate}:

\begin{equation}
P_{\text{active}} = \iint_\text{active area} \frac{1}{2}n_0\sqrt{\frac{\epsilon_0}{\mu_0 }}E^2(x,y,z=f) \,dx\,dy
\end{equation}

\noindent where $n_0$, $\epsilon_0$, $\mu_0$, and $E(x,y,z = f)$ are the refractive index of air, vacuum permittivity, permeability, and the lateral electric field intensity distribution at the focal length. The second metric to evaluate the microlens light collection is the concentration factor (CF), which is defined as \cite{bruschini2023challenges}:

\begin{equation}
  CF =  \frac{CE_\text{with microlens}}{CE_\text{without microlens}}
\end{equation}

\begin{figure}[ht!]
\centering

\includegraphics[width= 1\linewidth]{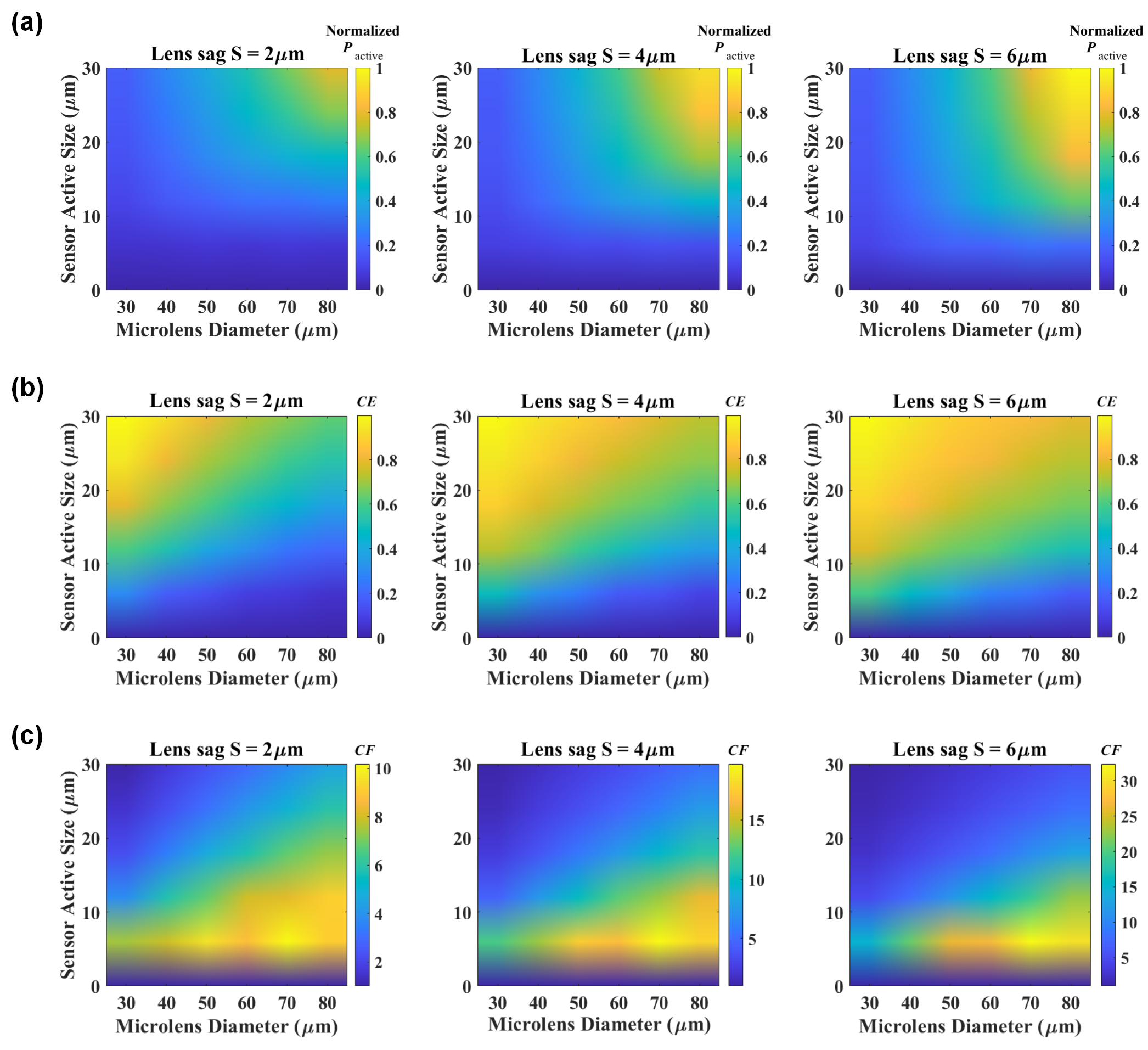}
\caption{Evaluation of light-collection performance. (a) Normalized power incident on sensor active area, (b) collection efficiency, and (c) concentration factor of Al$_2$O$_3$ spherical plano-convex microlenses with different lens sags in the MWIR, versus microlens diameter and sensor active size.}
\label{fig2}
\end{figure}

In \autoref{fig2}, we plot the change in these metrics for different sensor active areas, microlens diameters, and lens sags. We assume a square active area with side lengths equal to the sensor's active size. A schematic of pixel geometries is included in Section 5.2. In \autoref{fig2} (a), the power on sensor active area is normalized against the maximum value. The colorbar scales from $0$ to $1.0$, where $1.0$ represents the optimal geometric configuration for absolute signal yield, and all other values represent the exact relative percentage of signal degradation.

In the FDTD model, the incident wavefront illuminates the entire square unit cell ($D^2$). The light that hits the flat interstitial regions (the corners outside the circular lens perimeter) is not lost or absorbed; it propagates downward as unfocused light. We find that CE is maximized when the active size of the sensor matches the microlens diameter. As the sensor active area analytically expands to fill the entire square unit cell ($A_{\text{active}} \to A_{\text{pixel}}$), this unfocused interstitial light is captured alongside the central focal spot. At this limit of $100\%$ fill factor, $CE_{\text{with microlens}} = 1$ and $CE_{\text{without microlens}} = 1$, yielding a Concentration Factor ($CF$) of exactly $1.0$, mathematically demonstrating that the lens provides no net geometric advantage when the sensor already covers the entire pixel area. When real-world fabrication penalties (e.g., low transmittance) are applied to this $100\%$ fill factor regime, the lensed system performs worse than the bare sensor.

CF is the highest when the active size of the sensor is between 5 $\mu m$ and 10 $\mu m$. As the sensor size is reduced toward the dimensions of the focal spot (e.g., 5 to 10 µm), it progressively excludes the dark, low-intensity peripheral regions of the pixel. During this reduction, the baseline bare-pixel efficiency ($CE_{\text{without microlens}}$) shrinks rapidly while the lensed efficiency ($CE_{\text{with microlens}}$) remains relatively stable, driving the CF ratio upward. The CF reaches its absolute peak precisely when the sensor perimeter perfectly bounds the focal spot. At this optical sweet spot, the geometric advantage is maximized without sacrificing the collected energy. If the sensor is shrunk any further, its physical boundary begins cutting into the central concentrated beam. At that point, $CE_{\text{with microlens}}$ drops proportionally with the sensor area, perfectly canceling out any further geometric advantage and establishing the maximum theoretical CF limit shown in the colormaps.

With these simulations, we can design a microlens array with optimized performance based on the sensor array pitch and active area. These simulated trends are relevant not only to spintronic Poisson bolometers but also to other MWIR detector platforms with similarly constrained active areas.

We further evaluate the performance of the microlens array in the context of an infrared sensor. We examine its effect on the sensor's thermal sensitivity. In particular, we compute the noise-equivalent differential temperature (NEDT), a key figure of merit for characterizing the sensitivity of mid-wave infrared (MWIR) and long-wave infrared (LWIR) cameras \cite{rogalski2019infrared}. NEDT is defined as the temperature difference of a black body in the scene that produces a signal equal to the temporal noise of the camera. In other words, NEDT represents the minimum temperature difference that the camera can resolve. NEDT is calculated by dividing the temporal noise by the response per degree (responsivity), usually expressed in units of millikelvin. The NEDT measured at room temperature is defined as \cite{rogalski2019infrared, HowisNEDTonline}:

\begin{equation} 
NEDT = \langle \frac{ \displaystyle \sigma(T = 25^\circ\mathrm{C})} { \displaystyle \frac{\bar N(T = 30^\circ\mathrm{C}) - \bar N(T = 20^\circ\mathrm{C})}{10^\circ\mathrm{C}}} \rangle_{\text{array}}
\label{nedt}
\end{equation}

\noindent where $\langle \dots \rangle_{\text{array}}$ is the expected value notation for the spatial array average, and \(T\) is the temperature of the measurement. Because the counts are stochastic Poisson events, the number of counts registered by any single pixel fluctuates frame-to-frame, even when observing a stable calibration source. The variables $\bar N$ and $\sigma$ are strictly temporal statistics. $\bar N$ is the temporal mean (the expected number of counts for a specific pixel averaged over multiple frames). $\sigma$ is the temporal standard deviation of those counts for that same pixel. Because every pixel has a uniquely calculated NEDT, we must condense this NEDT matrix of a pixel array into a single representative figure of merit for the camera. We do this by taking the spatial average of all individual pixel NEDT values, i.e., $\langle \dots \rangle_{\text{array}}$.

The denominator in \cref{nedt} is a finite difference approximation of the continuous derivative $d\bar N/dT$ (the thermal responsivity of the system) evaluated at room temperature ($25^\circ\text{C}$). While taking a very small temperature differential (e.g., $1^\circ\text{C}$ or $0.1^\circ\text{C}$) would theoretically yield a stricter localized derivative, we adopted the widely established industry standard protocol for thermal imaging characterization (e.g., FLIR standard methodologies \cite{HowisNEDTonline}), which utilizes a $\Delta T$ of $10^\circ\text{C}$ (measuring at $20^\circ\text{C}$ and $30^\circ\text{C}$). For simulation purpose, we assume the response of the bolometer is highly linear across this specific $20^\circ\text{C}$ to $30^\circ\text{C}$ ambient regime, and this macroscopic finite difference ($\Delta \bar N / \Delta T$) equates to the continuous derivative ($d\bar N/dT$) without introducing non-linear approximation errors.



We model the response of a spintronic Poisson bolometer to incident light using a stochastic process governed by a Poisson random variable with mean \cite{leif2024thesis}:

\begin{equation} 
\mu = (\eta_{\text{yield}}*\lambda_\text{in}+\lambda_\text{dark})*t_\text{r}
\label{eq_sun1}
\end{equation}

\noindent where \(\eta_{\text{yield}}= \lambda_\text{ph} / \lambda_\text{in}\) is the empirical photon-to-transition coefficient, \(\lambda_\text{ph}\) is the detected count rate, \(\lambda_\text{in}\) is the incident photon rate, \(\lambda_\text{dark}\) is the dark count rate, and \(t_r\) is the detector response time. A larger empirical yield means that fewer incident photons, i.e., less incident optical power, are required to measurably increase the count rate above the thermal noise floor, which directly improves the responsivity and lowers the NEDT of the bolometer. 

The total detected counts \(N_\text{total}\) after some integration time \(t_\text{int}\) is:

\begin{equation} 
N_\text{total} = \sum_{i=1}^{t_\text{int}/t_\text{r}} N_i
\label{eq_sun2}
\end{equation}

\begin{equation} 
N_i =
\begin{cases} 
1 & \text{if } \text{Poisson}(\mu) \geq 1 \\
0 & \text{if } \text{Poisson}(\mu) = 0 
\end{cases}
\label{eq_sun3}
\end{equation}

\noindent where $N_i$ is the number of counts during each response time, and a maximum of one count can be detected during one response time window. The probability of a Poisson distribution is:

\begin{equation} 
P[\text{Poisson}(\mu) = n] = \frac{\mu^n}{n!}e^{-\mu}, n = 0,1,2,...
\label{eq_sun4}
\end{equation}

\noindent In \cref{eq_sun1}, the bracketed term represents the absolute transition rate (transitions per second). Multiplying this rate by the detector response time ($t_r$) yields $\mu$, the expected number of transitions within that time window. As a statistical average, $\mu$ is a continuous variable that can mathematically exceed $1.0$ under high illumination. Because $t_r$ represents the fundamental temporal resolving limit of the SPB readout circuit, the detector cannot distinguish multiple rapid transitions occurring within the same $t_r$ window. \autoref{eq_sun3} enforces this physical binary limit: if the underlying stochastic process yields one or more transitions ($Poisson(\mu) \ge 1$), the circuit registers a single digital count ($N_i = 1$).

To include the effect of the main lens, we use the radiative transfer model \cite{parr2005radiometry}:

\begin{equation} 
\Phi_\text{m} = L_\text{bb} \cdot \Omega_\text{m} \cdot A_\text{m}
\label{flux1}
\end{equation}

\begin{equation} 
\Phi_\text{SPB} = \Phi_\text{m} \cdot T_\text{m} \cdot CE
\label{flux2}
\end{equation}

\noindent where \(\Phi_\text{m}\) is the radiant photon flux incident on the microlens with units of \((\text{photons}/s)\), \(L_\text{bb}\) is the radiance from the blackbody source \((\text{photons}/s/m^2/sr)\), \(\Omega_\text{m}\) is the solid angle formed by a point on the microlens and the main lens, \(A_\text{m}\) is the area of the microlens,  \(\Phi_\text{SPB}\) is the photon flux incident on a spintronic Poisson bolometer, i.e., $\lambda_\text{in}$, \(T_\text{m}\) is the transmission of a microlens, and $CE$ is the collection efficiency of the microlens. A detailed discussion of the approximations made here is included in Section 3.2.

\begin{figure}[t!]
\centering
\includegraphics[width= 0.9\linewidth]{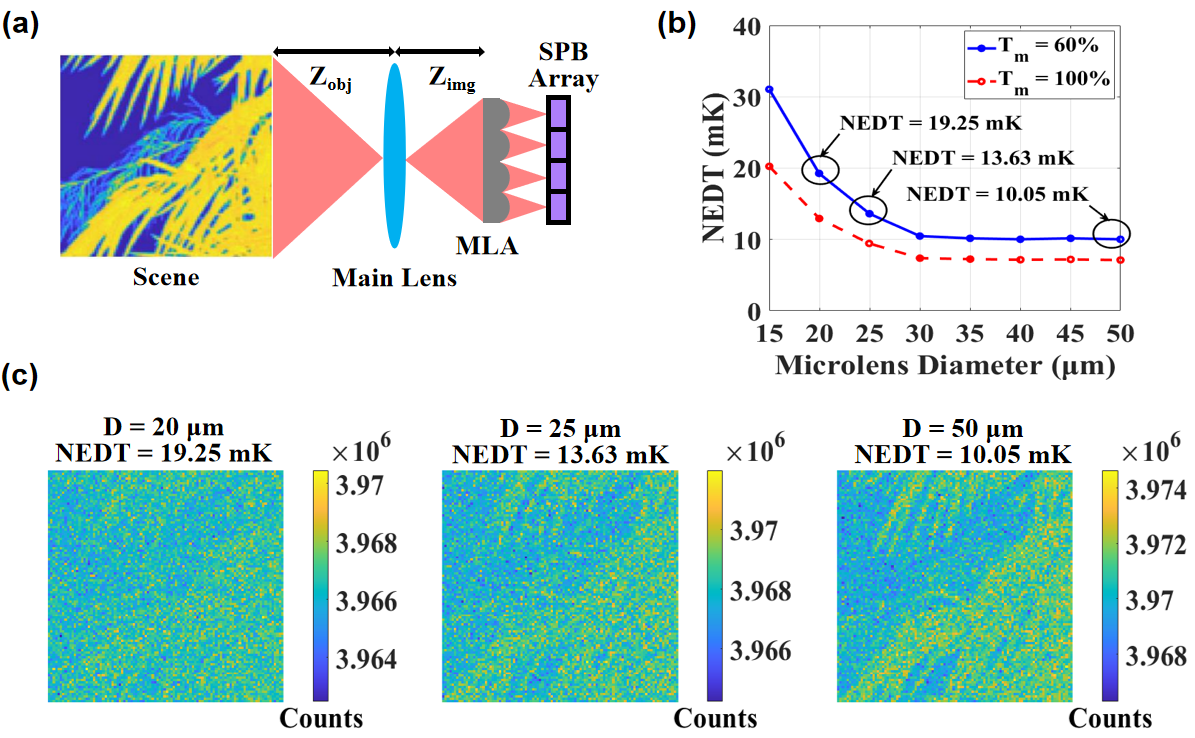}

\caption{Sensitivity improvement and full-image simulation of an SPB array with microlens arrays. (a) Imaging settings. (b) NEDT versus microlens diameter. \(T_\text{m}\)  is the transmission of the microlenses. (c) The effect of improved NEDT on a simulated image. A better contrast is shown with a smaller NEDT for different lens diameters.}
\label{fig4}
\end{figure}

We simulate the effect of an MLA on a spintronic Poisson bolometer when the MLA is placed at the image plane of the main lens. The object distance \(Z_\text{obj}\) between the blackbody and the main lens, and the image distance \(Z_\text{img}\) between the main lens and the MLA are determined by the thin lens equation \cite{BornWolf1999}:

\begin{equation} 
\frac{1}{Z_\text{obj}} +\frac{1}{Z_\text{img}} = \frac{1}{f_\text{main}}
\label{eq: thin_lens}
\end{equation}

\begin{table}[ht]
\caption{\bf Simulation Settings}

\centering
\begin{tabular}{ >{\centering\arraybackslash}p{6cm} c }
    
\toprule
\bf Parameters & \bf Value  \\      
\hline

Scene Temperature                                  & 300.15 K (27$^\circ\mathrm{C}$)\\
Scene Wavelength                                   & 3-5 $\mu m$ \\
Main lens Diameter and f/\#                                & 25.4 mm (f/0.5) \\
\(Z_\text{obj}\)                                       & 30 cm \\ 
\(Z_\text{img}\)                                       & 13.26 mm \\
$f_\text{main}$                                        & 12.7 mm\\
Microlens Sag \(S\)                                    & 6 $\mu m$ \\
Microlens Transmission \(T_\text{m}\)                 & 60\% and 100\% \\ 
Dark Count Rate \(\lambda_\text{dark}\) \cite{bauer2025exploiting}     & 3.9341 Mcps \\ 
$\eta_{\text{yield}}$ \cite{bauer2025exploiting}                                  & \(2.5 \times 10^{-8}\) \\
Integration Time \(t_\text{int}\)                         & 2 s \\ 
Response Time    \(t_\text{r}\) \cite{bauer2025exploiting}                          & 2 ns \\
Sensor Active Size \cite{bauer2025exploiting}                                       & 10 $\mu m$ \\

\bottomrule   
\end{tabular}
\label{ta:NEDT_sim_params}
\end{table}

\noindent where $f_\text{main}$ is the focal length of the main lens. The simulation settings are listed in \autoref{ta:NEDT_sim_params}. We show a schematic of the simulation in \autoref{fig4} (a), where one pixel is placed behind each microlens, and the detector array is placed at the focal length of the microlens array. The NEDT as a function of microlens diameter is shown in \autoref{fig4} (b). As defined in Table 1, the lens sag ($S$) in this simulation is fixed at $6\text{ }\mu\text{m}$. As the lens diameter ($D$) increases while the sag remains fixed, the lens geometry becomes increasingly flat, resulting in a longer focal length and a larger focused spot size (as demonstrated by the FWHM trends in \autoref{fig1.5}(h)). In the $D = 15$ to $25\text{ }\mu\text{m}$ regime, the focused spot is sufficiently tight to fall entirely within the fixed $10\text{ }\mu\text{m}$ active area, leading to a rapid improvement in sensitivity (lower NEDT). However, as $D$ exceeds $30\text{ }\mu\text{m}$, the focused beam waist inflates beyond the $10\text{ }\mu\text{m}$ boundary of the sensor. Consequently, any additional radiant power collected by the larger lens aperture is focused outside the active area. This geometric spillover causes the total collected power ($P_{\text{active}}$) to saturate, which mathematically results in the observed NEDT plateau. 

We note that the MLA transmission will also affect the NEDT. We also illustrate the effect of the improved NEDT on image quality using image simulations based on \cref{nedt,eq_sun1,eq_sun2,eq_sun3,eq_sun4,flux1,flux2} for the 3 microlens diameters labeled on \autoref{fig4}(b). The results are shown in \autoref{fig4}(c). As NEDT reduces, the image contrast improves, as expected by the improved sensitivity. Similarly, for emerging MWIR detector architectures where complex in-pixel electronics or engineering tradeoffs reduce fill factor and degrade optical coupling, integration with a microlens array can potentially enhance sensitivity and imaging performance \cite{pan2025design, bruschini2023challenges}.

We mentioned three distinct transmission values (80\%, 100\%, and 60\%) above. The 80\% transmittance is the pre-fabrication transmittance of a planar $\text{Al}_2 \text{O}_3$ substrate in the MWIR \cite{crystran_sapphire}. This value justifies our initial selection of $\text{Al}_2 \text{O}_3$ as a highly suitable optical material for this study. We use the perfectly lossless transmittance (100\%) for our baseline Collection Efficiency (CE) and Concentration Factor (CF) simulations, as well as our theoretical limit NEDT models. This allows us to isolate and evaluate the pure geometric and optical physics of the microlens architecture, independent of any material or manufacturing defects. It establishes the absolute theoretical ceiling of the design. The 60\% transmittance is the actual, experimentally measured transmittance of our fabricated microlens array using FTIR (\autoref{fig2}(d)). The FTIR measurement captures reflection at the top and bottom interfaces, absorption through the bulk material and scattering from the etched surface, but not the fill factor. The drop from 80\% to 60\% is primarily due to surface scattering, which is induced by increased surface roughness from the reactive ion etching (RIE) process. We apply this 60\% penalty strictly to our final, realistic NEDT models to demonstrate the practical, empirically bounded performance of the device as currently manufactured (\autoref{fig4}(b) and (c)).

\section{Discussions}
\subsection{Real-world CF limitation}
In the simulation of \autoref{fig2}, we assumed ideal, lossless optical transmission ($T_m=1$, subscript $m$ for ``microlens"). Real-world transmittance is dominated by unpredictable, fabrication-dependent artifacts such as increased surface roughness during the etching process and substrate qualities from the manufacturer. Attempting to parameterize these stochastic fabrication defects into the FDTD model would arbitrarily conflate the intrinsic optical capability of the lens geometry with extrinsic manufacturing limitations. By maintaining an idealized $T_ m=1$ baseline, readers can independently apply their own empirical transmission coefficients to our results to evaluate the real-world viability of these designs for their specific systems. For the fabrication in this work, the device transmittance reduced to 60\%, the real-world Concentration Factor (CF) must exceed 1.67 to yield a net positive signal gain over the bare sensor.
\subsection{Approximations for FDTD and Imaging simulations}

One objective of this FDTD study is to characterize the intrinsic, device-level optical properties of the integrated microlens, specifically isolating its diffraction-induced focal shift and inherent optical efficiency. To rigorously extract these inherent microlens parameters, it is necessary to decouple the microlens from the arbitrary numerical aperture (NA) and aberrations of a macroscopic optical system, i.e., the camera lens. Utilizing a normally incident plane wave to establish this fundamental baseline is a common methodology utilized in recent literature for FDTD simulations and experiments of microlenses \cite{kang2022substrate, bruschini2023challenges}.

For the simulation of \autoref{fig4}, a few approximations are made. The blackbody source is assumed to be an ideal Lambertian emitter, so photons are emitted diffusely into all directions. In addition, we assume that the main lens is lossless so that the radiance of the blackbody is conserved. To simplify the expression, we also assume that both the main lens and the microlens have small opening angles and that the Abbe sine condition is satisfied. A microlens array is typically used to direct more light onto the active area of a pixel when the incident light is a plane wave or has a small angular spread, as in the case of collimated laser illumination. For imaging applications, where light is usually scattered and incident from a wide range of angles, the focusing ability of each microlens becomes less effective. A conventional lens cannot focus highly divergent or strongly scattered light to a single point because the rays arrive over a broad angular distribution. Therefore, this treatment represents an upper-bound or idealized estimate of the sensitivity improvement provided by the microlens array.

\subsection{Power on sensor active area vs. CE}

\autoref{fig2}(a) maps the normalized power on sensor active area ($P_{\text{active}}$). This power is heavily dominated by the total geometric cross-section of the incident wavefront ($D^2$). In the top-right corner, both the microlens diameter ($D$) and the sensor active size ($S$) are at their maximum. This represents the largest possible physical area collecting the maximum absolute number of photons, naturally yielding the highest total power. \autoref{fig2}(b) maps Collection Efficiency ($CE$), a proportional metric, strictly defining the fraction of incident light captured ($P_{\text{active}} / P_{\text{total}}$). In the top-left corner, the microlens diameter ($D$) is small, meaning the absolute incident power is very low. However, in this same quadrant, the sensor active size ($W$) is relatively large, meaning $W$ approaches $D$. Because the sensor area expands to cover nearly the entire unit cell (approaching a 100\% fill factor), it successfully captures nearly all of that small amount of incident light. Capturing 100\% of a small incident power still yields a maximum Collection Efficiency of $1.0$. In summary, \autoref{fig2}(a) shows where the most total energy is collected, while \autoref{fig2}(b) shows where the system is most spatially efficient.

\section{Conclusion}

In conclusion, we used FDTD simulations to study how key design parameters govern the optical performance of MWIR microlens arrays. Using these models, we studied the collection efficiency and concentration factor as a function of microlens diameter and sensor active size. Furthermore, we fabricated a microlens array sample to demonstrate that the chosen geometrical parameters are realistic and compatible with the fabrication process. Based on the simulated microlens performance, we find that NEDT can be reduced and image contrast improved in MWIR imaging with spintronic Poisson bolometers. This enhanced performance may enable the use of spintronic Poisson bolometer arrays in neuromorphic computing and uncooled MWIR imaging applications \cite{mousa2025neural}. Our work presents the first simulation of microlens design for spintronic Poisson bolometer arrays, bridging microphotonics, spintronics, and thermal imaging. Beyond the specific detector demonstrated here, the simulations establish practical MWIR microlens-array scaling trends that can guide optical concentration design for other detector arrays with limited active area or fill factor.

Currently, our custom SPB arrays are wire-bonded directly to printed circuit boards (PCBs) for fundamental electrical characterization. Integrating these existing spintronic chips with a precisely aligned, microscopic $\text{Al}_2 \text{O}_3$ lens array requires advanced heterogeneous integration techniques (such as sub-micron flip-chip bump bonding \cite{xu2013wafer}), which is the critical next step for future experimental work. By experimentally demonstrating that the exact microlens geometries (diameters and sags) and MWIR transmission profiles used in our simulations can be reliably manufactured, we prove that our theoretical concentration factors are built on practically attainable manufacturing parameters. As a result, our current performance enhancements are derived from a physically validated optical model, and the ultimate milestone for this technology is the experimental demonstration of a fully integrated MLA-SPB system.

\section{Methods}
\subsection{FDTD settings}

\begin{table}[ht]
\centering
\caption{FDTD Simulation Settings}
\begin{tabular}{l p{10cm}}
\toprule
\textbf{Parameter} & \textbf{Value / Setting} \\
\midrule
\textbf{FDTD region} & $[x] = [-D/2, +D/2]$, $[y] = [-D/2, +D/2]$, $z = [-10, +100]$ $\mu m$ \\
 & ($D$: microlens diameter; equal to pixel width) \\
\addlinespace
\textbf{Grid size} & Ranges from $\approx 0.238$ $\mu m$ to $0.398$ $\mu m$ as $D$ varies from $30$ to $80$ $\mu m$. \\
\addlinespace
\textbf{Substrate} ($\text{Al}_2 \text{O}_3$, $n = 1.685$) & $[x] = [y] = [-D/2, +D/2]$, $z = [-10, 0]$ $\mu m$ \\
\addlinespace
\textbf{Microlens} ($\text{Al}_2 \text{O}_3$, $n = 1.685$) & $[x] = [y] = [-D/2, +D/2]$, $z = [0, +S]$ ($S$: lens sag) \\
\addlinespace
\textbf{Source} & Plane wave located at $z = -2$ $\mu m$ (inside the substrate); \\
 & Wavelength $\lambda = 3-5$ $\mu m$; linear polarization. \\
\bottomrule
\end{tabular}
\label{tab:fdtd_params}
\end{table}

The optical axis of the system lies along the z-axis, the z-min and z-max boundaries were terminated using an 8-layer stretched coordinate Perfectly Matched Layer (PML). To rigorously model the close-packed array architecture, the transverse boundaries (x-axis and y-axis) were set to Periodic boundary conditions, defining the $D \times D$ square unit cell.

The refractive index of $\text{Al}_2 \text{O}_3$ we used for all the FDTD simulations was $n=1.685$. In our FDTD models, the incident plane wave was a broadband MWIR pulse spanning 3 µm to 5 µm. The complex refractive index of the $\text{Al}_2 \text{O}_3$ microlenses was extracted directly from the Palik database (Handbook of Optical Constants of Solids) in Lumerical FDTD. Because $\text{Al}_2 \text{O}_3$ exhibits relatively low dispersion across this specific MWIR window, the FDTD solver's material fitting algorithm optimized computational efficiency and stability by utilizing a constant refractive index of 1.685.

Because our spherical microlenses are perfectly rotationally symmetric, the choice of linear polarization orientation does not alter the total optical power integrated over the sensor's active area. Physically, unpolarized light can be modeled as the incoherent superposition of two orthogonal linearly polarized states. Due to the high structural symmetry of both the microlens and the square sensor active area, the focusing behavior and collection efficiency for $x$-polarized and $y$-polarized light are identical.

\subsection{System configuration and pixel geometries}

We chose to face the curved lens towards the sensor rather than towards the incident light solely due to fabrication concerns. If the flat surface is faced towards the sensor, then, without wafer bonding, we would have to deposit the microlens directly onto the sensor using sputtering. However, there is a limitation on the maximum thickness allowed to be deposited for common deposition tools (300 nm in our case). While the focal length is at least microns, it is highly impractical to deposit a material directly thick enough to locate the sensor on the focal plane. With wafer bonding, it is again not feasible to find a substrate as thin as hundreds of microns to locate the sensor on the focal plane. Common substrates are at least a millimeter thick, and the focal plane will be located inside the substrate. As a result, we had to face the microlens towards the sensor to adapt for wafer bonding for the integration of the microlens and the sensor.

\begin{figure}[h!]
\centering
\includegraphics[width= 0.3\linewidth]{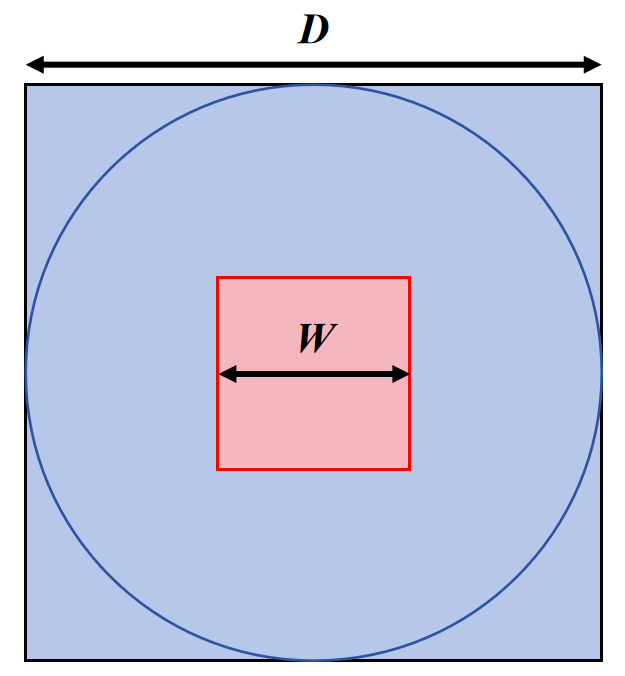}
\caption{Geometrical layout of a pixel.}
\label{PixGeo}
\end{figure}

We assume a square pixel array. The pixel is a square with side length $D$. The lenses are close-packed in this square array. Therefore, the microlens has a circular base with a diameter equal to the pixel width ($D$). The sensor itself is square, with side width ($W$) labeled on \autoref{fig2} as ``Sensor Active Size".

\subsection{Mechanical focal length}
Mathematically, our extracted mechanical focal length (MFL) $f_m$ relates to the theoretical vertex measurement simply by adding the lens sag ($f_m = EFL + S$). Because our lens sag ($S = 2 \text{ to } 6 \ \mu m$) is small compared to the EFL (which is larger than $20 \ \mu m$), this $z$-axis shift in the reference plane does not change the conclusion drastically. For wafer bonding, the bulk planar substrate serves as the primary mechanical datum (reference plane) for depositing spacer layers. Measuring from this flat interface is more practical for experimental device integration. 

\subsection{Calculation of CE}
The Collection Efficiency (CE) was calculated using the intensity distribution at the focal plane, as shown in \autoref{fig1.5}(d). To obtain the cross section intensity distribution at the focal plane, we used a rigorous two-step monitor configuration. First, we record 2D longitudinal E-field distribution along the optical axis (as shown in \autoref{fig1.5}(c)), allowing us to precisely extract the focal length ($f$) for each lens geometry. Second, a 2D transverse power monitor was placed exactly at this extracted focal plane to capture the cross-sectional E-field intensity (as shown in \autoref{fig1.5}(d)). To calculate the CE, we integrated the normalized power over the spatial bounds of the sensor active area ($W \times W$).

%
%

\ack{We thank Professor Xianfan Xu for providing access to the MWIR Fourier Transform Infrared (FTIR) Microscopy instrumentation and Yikang Chen for guiding the transmission spectrum measurement. We thank the anonymous reviewers for their insightful comments and constructive suggestions.}

\funding{This work is partially supported by the Elmore Chaired Professorship of Purdue University.}





\bibliographystyle{vancouver}

\bibliography{reference}

\end{document}